\documentclass{IEEEtran}
\usepackage{cite}
\usepackage{amsmath,amssymb,amsfonts}
\usepackage{graphicx}
\usepackage{bm}
\usepackage{textcomp}
\usepackage{fancyhdr}
\usepackage{booktabs}
\usepackage{color}
\usepackage[table,xcdraw]{xcolor}
\usepackage{graphicx,epstopdf}
\usepackage{subfig} 
\def\BibTeX{{\rm B\kern-.05em{\sc i\kern-.025em b}\kern-.08em
    T\kern-.1667em\lower.7ex\hbox{E}\kern-.125emX}}

\fancypagestyle{plain}{
	\fancyhf{} 
	\fancyfoot[C]{\textbf{\thepage}} 
	
	}

\makeatletter
\long\def\@makecaption#1#2{
	\vskip\abovecaptionskip
	\sbox\@tempboxa{#1. #2}
	\ifdim \wd\@tempboxa >\hsize
	#1. #2\par
	\else
	\global \@minipagefalse
	\hb@xt@\hsize{\hfil\box\@tempboxa\hfil}
	\fi
	\vskip\belowcaptionskip}
\makeatother

\begin{document}

\onecolumn
\newpage
\thispagestyle{empty}

\textbf{Copyright:}
\copyright 2019 IEEE. Personal use of this material is permitted.  Permission from IEEE must be obtained for all other uses, in any current or future media, including reprinting/republishing this material for advertising or promotional purposes, creating new collective works, for resale or redistribution to servers or lists, or reuse of any copyrighted component of this work in other works.\\

\textbf{Disclaimer:} This work has been published in \textit{IEEE Microwave and Wireless Components Letters}. \\

Citation information: DOI 10.1109/LMWC.2019.2939553

\newpage

\twocolumn

\setcounter{page}{1}

\title{
Generalized Director Approach for Liquid-Crystal-Based Reconfigurable RF Devices}
\author{Antonio Alex-Amor, \'{A}ngel Palomares-Caballero,  Antonio Palomares, Adri\'{a}n Tamayo-Dom\'{i}nguez,  \\Jos\'{e} M. Fern\'{a}ndez-Gonz\'{a}lez, \textit{Senior Member, IEEE}, Pablo Padilla
\thanks{Manuscript submitted August 8, 2019; accepted September 2, 2019. This work was supported in part by the Spanish Research and Development National Program under Projects TIN2016-75097-P,  TEC2017-85529-C3-1-R and RTI2018-102002-A-I00. }

\thanks{A. Alex-Amor and \'{A}. Palomares-Caballero are with the Departamento de Lenguajes y Ciencias de la Computaci\'{o}n, Universidad de M\'{a}laga, 29071 M\'{a}laga,  Spain. (e-mail: aalex@lcc.uma.es; angelpc@uma.es)}
\thanks{A. Alex-Amor, A. Tamayo-Dom\'{i}nguez  J. M. Fern\'{a}ndez-Gonz\'{a}lez are with the Information Processing and Telecommunications Center, Universidad Polit\'{e}cnica de Madrid, 28040 Madrid, Spain (e-mail: aalex@gr.ssr.upm.es; a.tamayo@upm.es;  jmfdez@gr.ssr.upm.es)}
\thanks{A. Palomares is with the Departamento de Matem\'{a}tica Aplicada, Universidad de Granada, 18071 Granada, Spain (e-mail: anpalom@ugr.es)}
\thanks{P. Padilla is with the Departamento de Teor\'{i}a de la Se\~{n}al, Telem\'{a}tica y Comunicaciones, Universidad de Granada, 18071 Granada, Spain (e-mail: pablopadilla@ugr.es)}
}

\maketitle

\begin{abstract}
	This letter presents a closed-form integral-equation formulation that models the director tilting in nematic liquid crystals. The proposed formulation is computationally efficient compared to numerical methods and provides a physical insight on the matter. Our previous work is generalized from the one-constant approach to all the possible cases that cover the elastic constants. These constants determine in great extent the electrical properties of the liquid crystal, and subsequently, the response of the reconfigurable devices that take use of it. Therefore, a more precise but still simple modeling of their influence is pursued here. Simulations show good agreement with numerical implementations. In comparison, the error in the estimation is considerably reduced when using the integral-equation formulation, specially as the polarization voltage or the dielectric anisotropy decreases. 

\end{abstract}


\section{Introduction}
\IEEEPARstart{R}{econfigurable} radiofrequency (RF) devices allow us to control the radiation characteristics of a certain system. Tunable materials, such as liquid crystals (LC), are being massively applied in future millimeter-wave communication systems due to their promising capabilities \cite{review}. Specifically, nematic liquid crystals \cite{lc} are of special interest in microwave and millimeter-wave frequency ranges. In this phase, LC molecules orient in the same average direction, so the anisotropy in the material can be studied with a set of local vectors called directors.

The elastic properties related to liquid crystals are described by means of three parameters: $k_{11}$, $k_{22}$ and $k_{33}$, known as the Frank elastic constants \cite{elastic_constants}. These constants are associated to splay, twist and bend deformations \cite{splay_twist_bend}, respectively, and describe how the liquid crystal is distorted by the presence of an external electric $\mathbf{E}$ or magnetic $\mathbf{H}$ field. Thus, the design of liquid crystal-based reconfigurable devices, such as microwave phase shifters \cite{mwcl}, filters \cite{mwcl2}, polarizers \cite{polarizer} or antennas \cite{maci,gerardo,antenna_new}, is fully dependent on the value of the elastic constants. 

\begin{figure}[t]
	\vspace*{-0.3cm}
	\hspace*{-0.2cm}
	\subfloat[$V=0$]{
		\label{lc_v0v3}
		\includegraphics[width=0.242\textwidth]{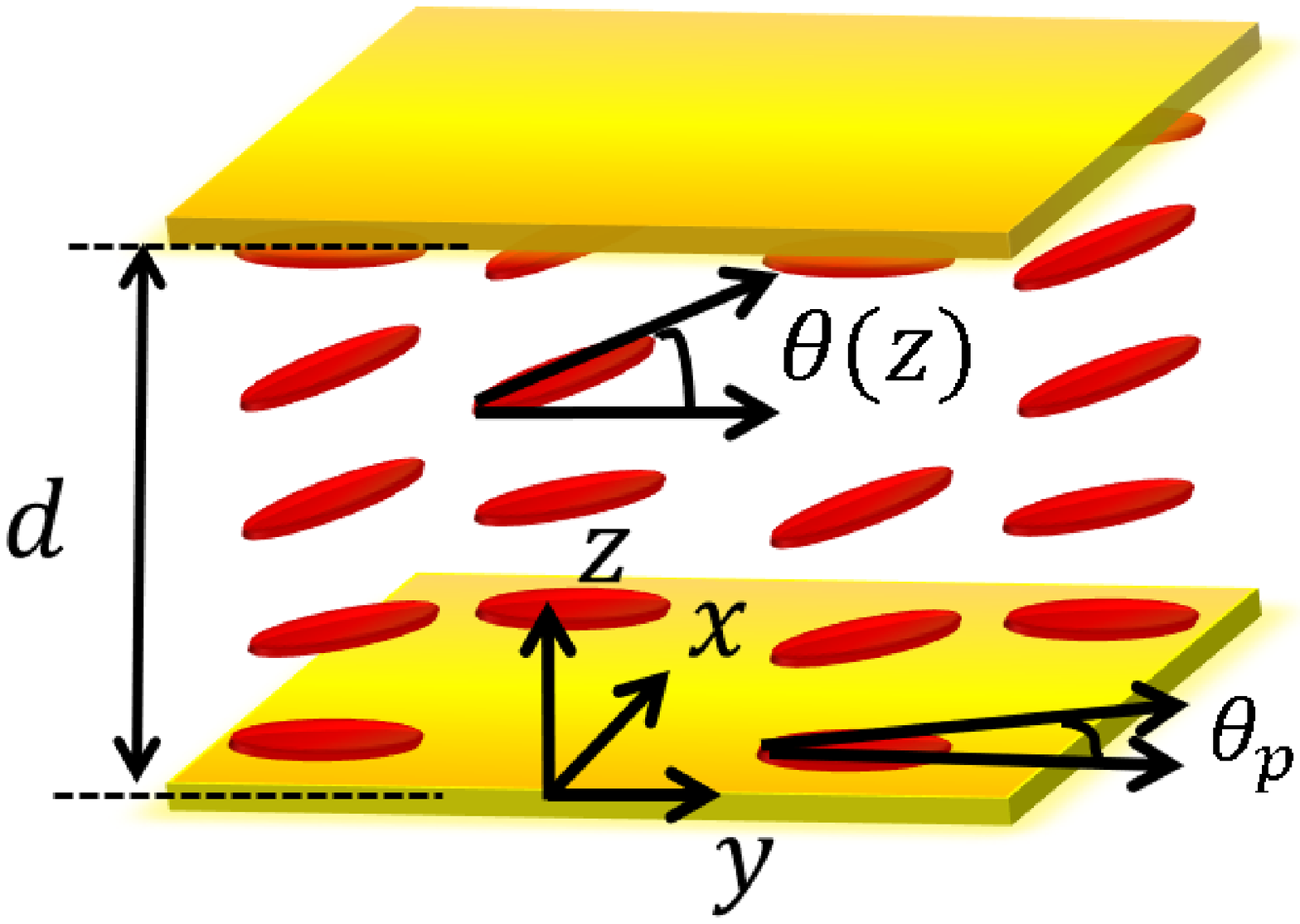}}
	\hspace*{-0.25cm}
	\subfloat[High voltage]{
		\label{lc_valtov3}
		\includegraphics[width=0.235\textwidth]{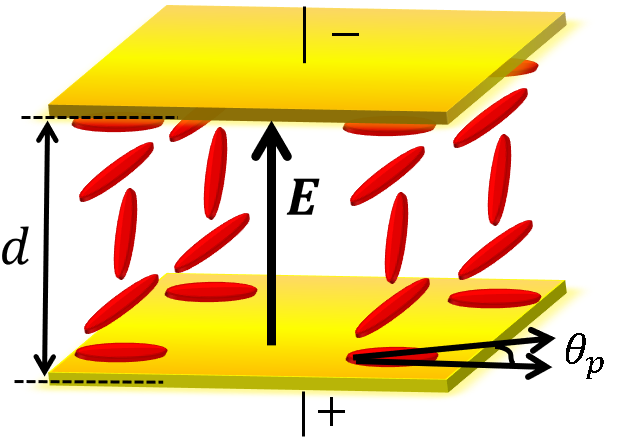}}
	\caption{\small Director tilting in a nematic liquid crystal with splay geometry when a null (a) and a high-intensity (b) electric field are applied.  }
	\label{lc}
\end{figure}

The splay geometry fits most of the practical implementations involving microwave devices. As depicted in Fig. 1, the liquid crystal is polarized by applying a potential difference between the two metallic plates where it is confined. This induces an electric field $\mathbf{E}$ that is perpendicular to the metallic plates (no fringing field is assumed) and forces the directors to orient parallel to it, as shown in Fig. \ref{lc}(b).

Modeling the electrical behavior of a nematic liquid crystal is a complex task: the governing second-order differential equation is highly nonlinear and lacks an analytical solution. The small-angle approximation provides a poor analytical estimation to the problem \cite{lc} that cannot be used in general for practical purposes. Thus, we are generally forced to perform numerical estimations (FEM, FDM, Runge-Kutta, etc.)\cite{mtt_rjames}, \cite{numerical} at the expense of losing the physical insight on the properties of the material. In \cite{Alex}, it is recently proposed an analytical approach that takes into account these considerations. However, this approach is restricted to the case $k_{11}=k_{22}=k_{33}$, commonly known in the literature as the one-constant approximation \cite{lc}. In this paper, we derive a closed-form integral equation that generalizes the modeling of the LC for all the possible values of the Frank elastic constants. This closed-form equation drastically reduces the computational complexity of numerically solving the differential equation and provides more accurate results compared to \cite{Alex}. Furthermore, it gives a useful insight on the physics of  microwave liquid-crystal-based reconfigurable devices.

The paper is organized as follows. Section II presents the derivation of the closed-form expression. Section III presents the simulation results, which are validated with numerical approaches and compared with \cite{Alex}. Finally, conclusions are drawn in Section IV.

\vspace*{-0.2cm}\section{Integral-Equation Formulation}
The differential equation that describes the orientation of directors $\theta(z)$ in a nematic liquid crystal was presented in \cite{lc}:
\begin{equation}\label{original} 
\begin{split}
( k_{11} &\cos^2   \theta +k_{33} \sin^2  \theta) \frac{d^2 \theta}{dz^2}\\ +
&\left( k_{33}- k_{11}  \right) \left( \frac{d \theta}{dz} \right)^2 \sin \theta \cos \theta +
\beta \sin \theta \cos \theta  =0
\end{split}
\end{equation}
subject to $\theta(0)=\theta(d)=\theta_p$, where $d$ is the thickness of the liquid crystal and $\theta_p$ is the pretilt angle. The metallic plates physically impose $\theta_p=0$ in the LC cell \cite{lc}. However, $\theta_p$ is slightly different from zero in real implementations due to the manufacturing process \cite{pretilt1,pretilt2}.  In \eqref{original}, $k_{11}$ and $k_{33}$ are the splay and bend elastic constants,  $\beta=\varepsilon_0 \Delta \varepsilon E^2$, $\Delta \varepsilon$ is the LC dielectric anisotropy, and $E=V/d$ is the module of the uniform electric field. As widely discussed in \cite{Alex}, the contribution of the term $\left( k_{33}- k_{11}  \right) \left( \frac{d \theta}{dz} \right)^2 \sin \theta \cos \theta$ to equation \eqref{original} is negligible and can be eliminated from the equation. Thus, the differential equation is simplified into 
\begin{equation}\label{simplified} 
 \frac{d^2 \theta}{dz^2}=\frac{-\beta \sin \theta \cos \theta }{ k_{11} \cos^2   \theta +k_{33} \sin^2  \theta}
\end{equation}
Note that, unfortunately, both \eqref{original} and \eqref{simplified} lack an analytical solution. However, a particular analytical approach can be derived from \eqref{simplified} when $k_{11}=k_{33}$. This fact was already exploited in \cite{Alex}. However, the accuracy of the analytical approach reduces as $k_{33}$ moves away from $k_{11}$. In this section, we derive a closed-form expression that considers all possible values of $k_{33}$ and $k_{11}$. Initially, we consider $k_{11} \neq k_{33}$ and then, we study the case $k_{11} = k_{33}$. 

Applying the change of variable
$u(\theta(z))=\frac{d\theta}{dz}(z)$ and using the chain rule, we
have $\frac{d^2\theta}{dz^2}=u \frac{d u}{dz}$. Then, the
integration of both sides of equation \eqref{simplified} leads to
\begin{equation} \label{change2}
	u^2=\left( \frac{\beta}{k_{11}-k_{33}}\right) 
	\ln \bigg( \big(k_{11}-k_{33}\big) \cos  \big(2\theta\big) +k_{11}+k_{33} \bigg) +c_0
\end{equation}
where $c_0$ is a constant of integration. To simplify the notation, we write $k_{11}+k_{33}=k^{+}$. If the problem has a unique
solution, it can be proved that $\frac{d\theta}{dz}(\frac{d}{2})=0$. We use this fact to clear the constant of integration $c_0$ from \eqref{change2}, leading to
\begin{equation} \label{change4}
\left( \frac{d\theta}{dz}\right)^2 = \left( \frac{\beta}{k_{11}-k_{33}}\right) 
\ln\left( \frac{(k_{11}-k_{33}) \cos  \big(2\theta(z)\big) + k^{+}}{(k_{11}-k_{33}) \cos  \big(2\theta(\frac{d}{2})\big) +k^{+}} \right)
\end{equation}
When $\theta$ is an increasing function, $\frac{d\theta}{dz}$ is positive, and we can isolate it taking square roots (the case $\theta$ decreasing is studied later). By integrating both sides of the equation between 0 and $z$ and noting that in $\int_0^z \frac{d\theta}{dz} dz= \theta(z)-\theta(0)$, $\theta(0)=\theta_p$, the closed-form integral equation is expressed as
\begin{equation} \label{k11mayor}
\begin{split}
\theta(z)=&\theta_p+ \int_0^z\left[ \left( \frac{\beta}{k_{11}-k_{33}}\right) \right.   \\
& \left. \times \ln\left( \frac{(k_{11}-k_{33}) \cos  \big(2\theta(z)\big) +k^{+}}{(k_{11}-k_{33}) \cos \big( 2\theta(\frac{d}{2})\big) +k^{+}} \right)\right]^{1/2} dz
\end{split}
\end{equation}
Note that this expression is only valid in the interval $0\leq z \leq d/2 $. If we want to move the constant outside the integral, we have to consider the cases $k_{11} < k_{33}$ and $k_{11} > k_{33}$. 

For the one-constant case $k_{11}=k_{33}=k$, we particularize \eqref{simplified} and we have
\begin{equation}\label{simplified2} 
\frac{d^2 \theta}{dz^2}=-\frac{\beta}{k} \sin \theta \cos \theta 
\end{equation}
 After applying the change of variable $u=\frac{d\theta}{dz}$ and integrating both sides of \eqref{simplified2}, the differential equation is rewritten as
\begin{equation}\label{change_simplified2} 
\frac{d\theta}{dz}=\left[ \frac{\beta}{k} \cos^2\big(\theta\big)+c_0 \right]^{1/2}
\end{equation}
To clear the constant of integration $c_0$, we newly use $\frac{d\theta}{dz}(\frac{d}{2})=0$. Rearranging terms, the integral equation that contemplates the case $k_{11}=k_{33}=k$ is expressed as
\begin{equation} \label{k33igual}
\begin{split}
\theta(z)=\theta_p+\left( \frac{\beta}{2k}\right)^{1/2} 
\int_0^z \bigg[\cos\big(2\theta(z)\big)- \cos\big(2\theta (\tfrac{d}{2}) \big)  \bigg]^{1/2} 
 dz
\end{split}
\end{equation}

The symmetry of the solutions obtained from \eqref{k11mayor} and \eqref{k33igual} is used for extending them to the interval $d/2\leq z \leq d$. Thus, the solutions for the latter interval are just reflections of $\theta(z)$ with respect to $z=d/2$. See the reflection axis placed at $z=100 \,\mu$m in Fig. \ref{2D} (yellow line) and the symmetry of $\theta(z)$.

Solving equations \eqref{k11mayor} and \eqref{k33igual} requires a previous estimation of $\theta (d/2)$. A good starting point is the value provided by \cite{Alex}, which is typically near $\pi/2$ (90$^\mathrm{o}$) for high polarization voltages. This value can be improved by using an iterative method. In order to ensure the convergence, the robust bisection method is utilized for this purpose. Finally, all integrals are computed with the trapezoidal method.



\section{Results}

In this section, we test the solutions obtained from the integral equations for all the parameters involved on the problem ($k_{11}, k_{33}, V, \Delta \varepsilon$). We validate the results with numerical estimations and show that the error in our proposed approach is considerably reduced compared to \cite{Alex}. In order to point out the implications of Section II in practical microwave designs, we will use as a reference the values measured in Merck E7 LC \cite{mtt}: $k_{11}=11.7$ pN, $k_{33}=17.1$ pN, $\Delta \varepsilon=0.47$, $\varepsilon_{\perp}=2.7$. The pretilt, which usually ranges between 0$^\mathrm{o}$ and 10$^\mathrm{o}$ \cite{lc}, is assumed to be 4$^\mathrm{o}$ in this work. Note that higher $\theta_p$ values will imply higher tilt angles and lower threshold voltages \cite{pretilt3}.

\begin{figure}[t]
	\vspace*{-0.2cm}
	\hspace*{-0cm}
	\includegraphics[width=0.42 \textwidth]{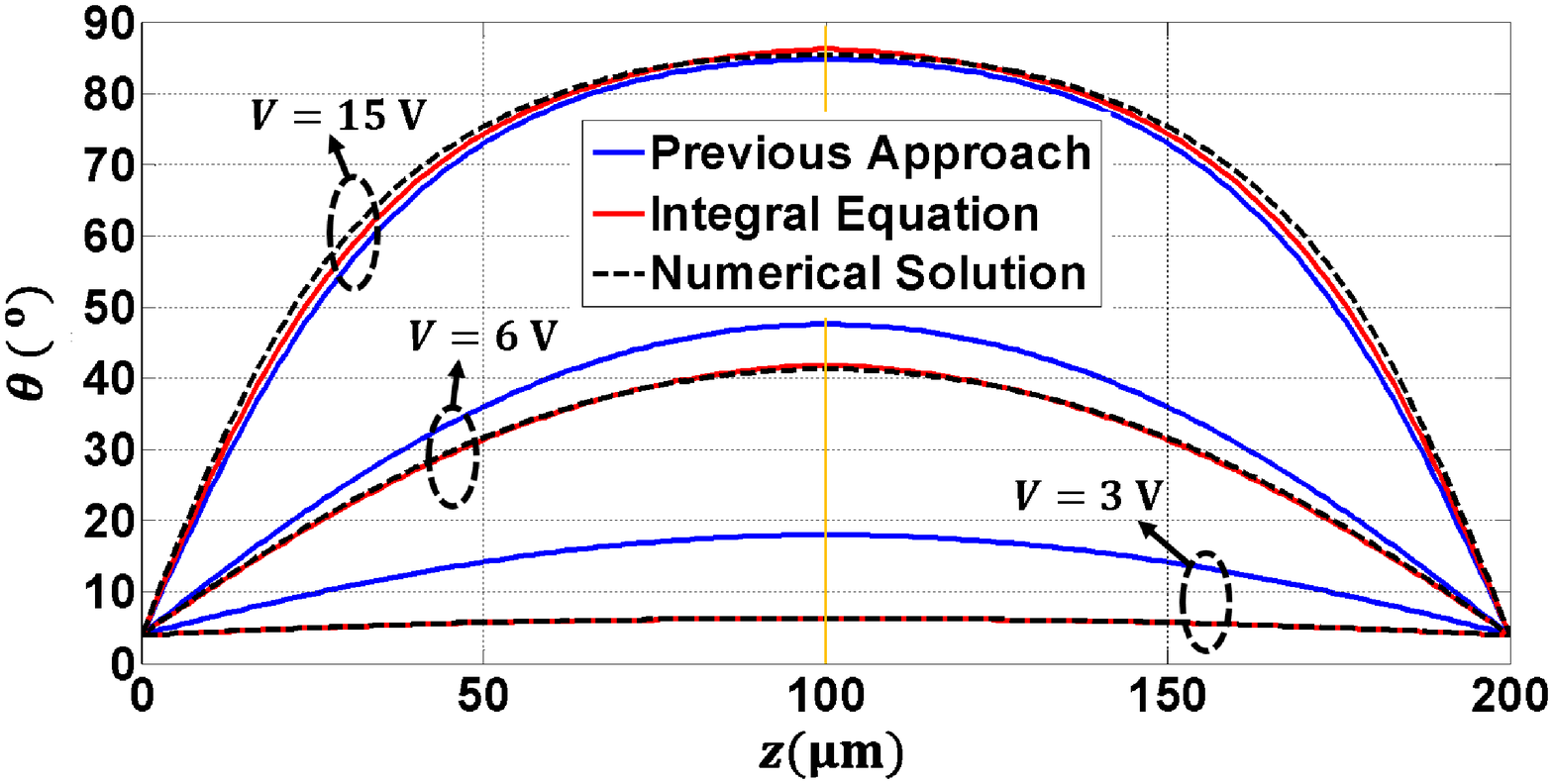}
	\centering
	\caption{\small Director tilting in a Merck E7 LC for three different voltages ($k_{11}=11.7$ pN, $k_{33}=17.1$ pN, $\Delta \varepsilon=0.47$ and $\theta_p=4^\mathrm{o}$).}
	\label{2D}
\end{figure}

Fig. \ref{2D} presents the director tilt angle $\theta(z)$ in a 200-$\mu$m Merck E7 LC for three different polarization voltages. The agreement between our formulation and numerical estimations is very good for the three cases under study. However, the error increases when using the approach of \cite{Alex} (blue curves) as the applied voltage decreases. As shown in the figure, the directors tend to orient perpendicular to the metallic plates (90$^\mathrm{o}$) as the applied voltage is bigger. In addition, the dependence of the electric field on the governing differential equation \eqref{original} is quadratic ($\beta=\varepsilon_0 \Delta \varepsilon E^2$). Thus, slight changes in the applied voltage cause appreciable variations in the director tilting, and subsequently, in the anisotropic response of the LC. 

\begin{figure}[t]
	\vspace*{-0.55cm}
	\hspace*{-0.2cm}
	\centering
	\subfloat[\hspace*{-0.7cm}]{
		\label{previous_approach1}
		\includegraphics[width=0.23\textwidth]{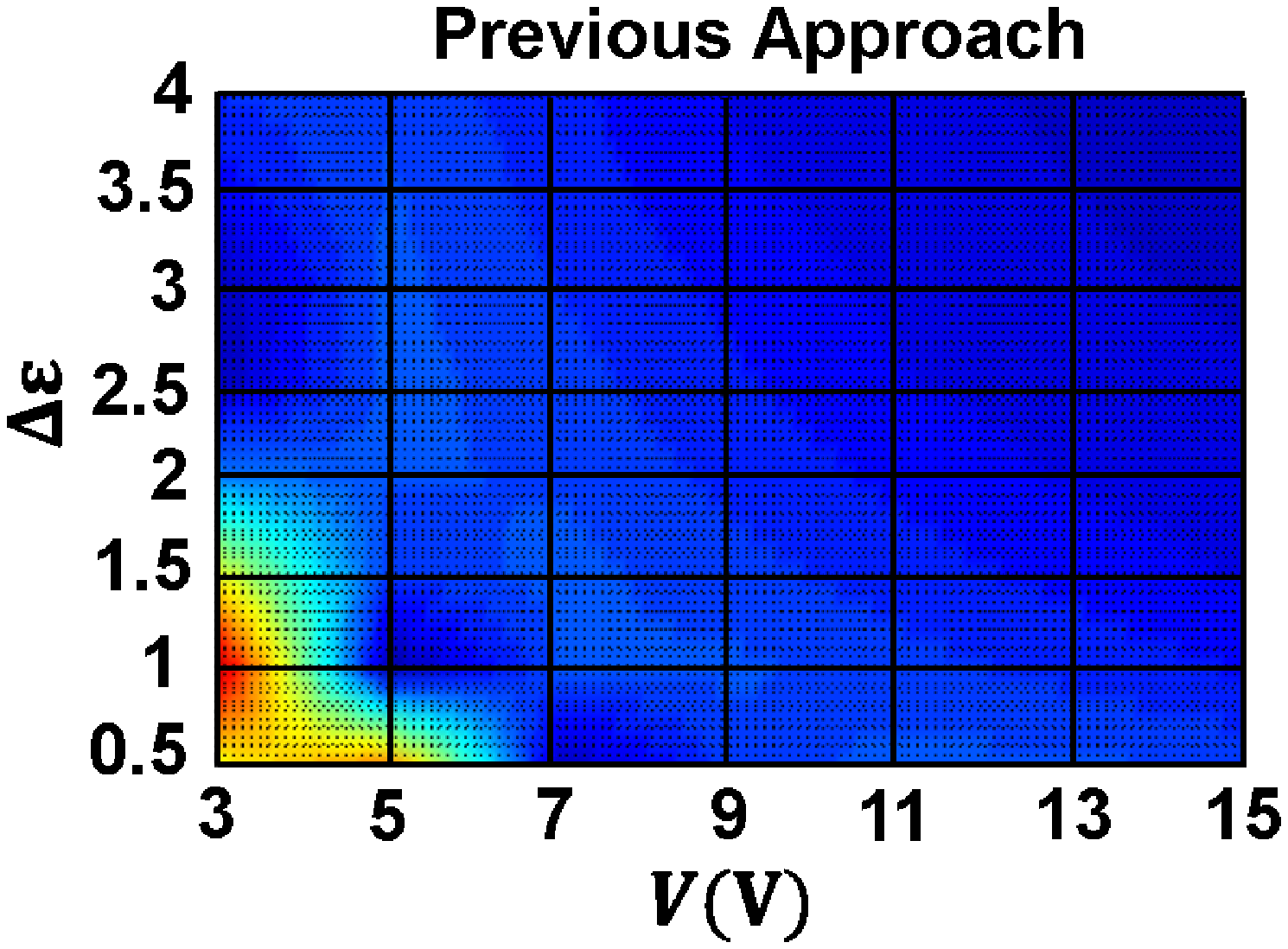}}
	\hspace*{-0.2cm}
	\subfloat[\hspace*{-0.7cm}]{
		\label{integral_approach1}
		\includegraphics[width=0.23\textwidth]{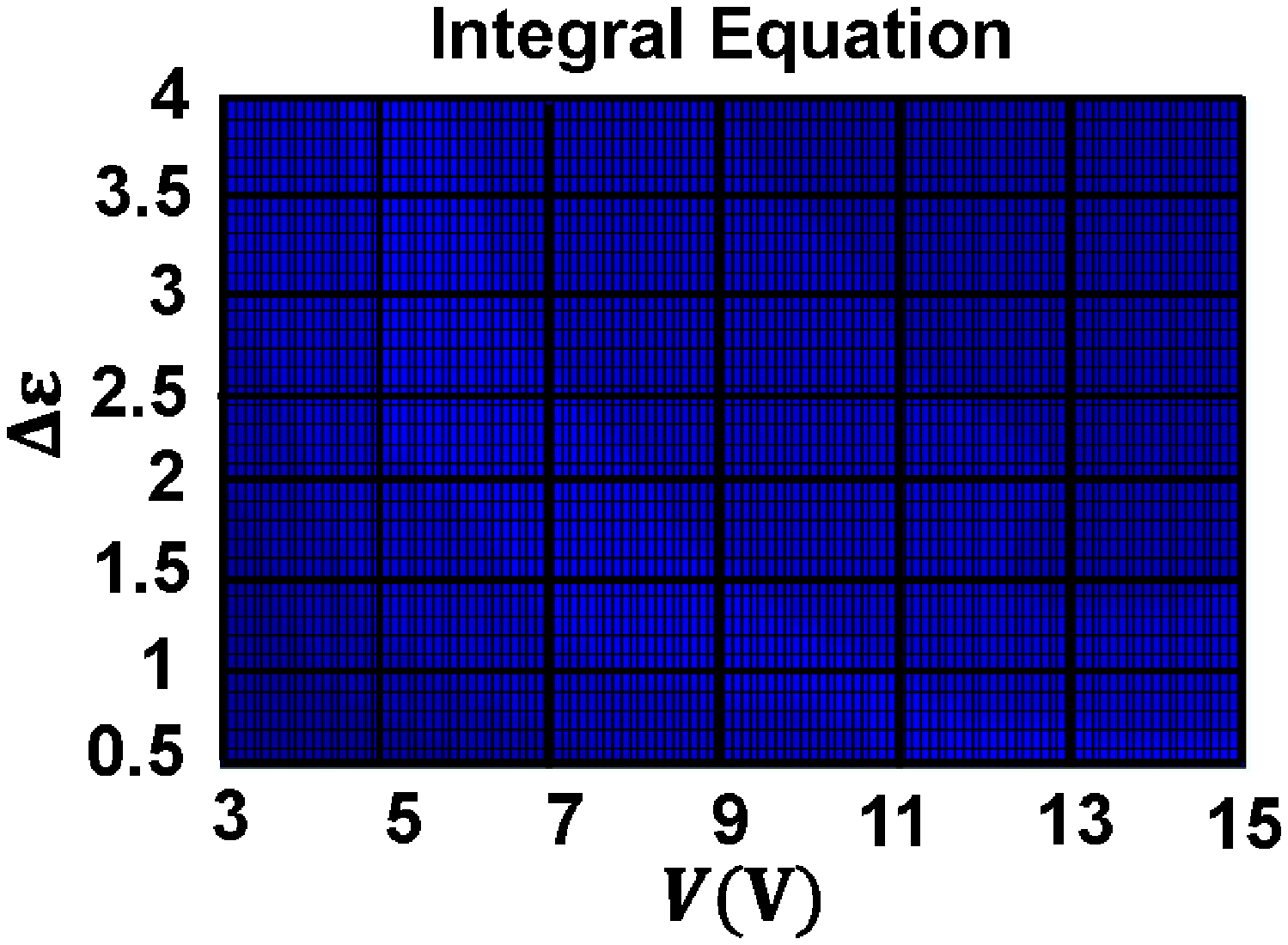}}
	\hspace*{-0.50cm}
	\includegraphics[width=0.058\textwidth]{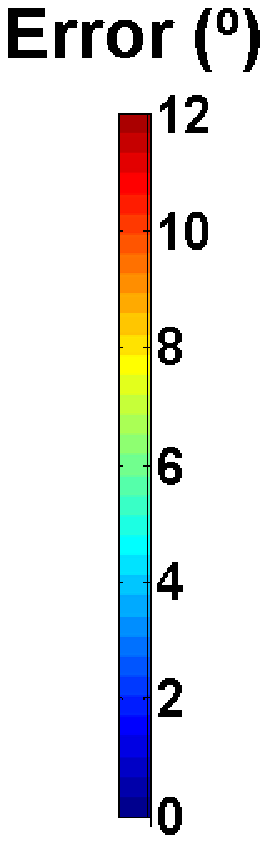}
	\caption{\small Mean absolute error as a function of $V$ and $\Delta\varepsilon$ obtained with the approach of \cite{Alex} (a) and the integral-equation formulation (b) ($k_{11}=11.7$ pN, $k_{33}=17.1$ pN, $d=200 \, \mu$m, and $\theta_p=4^\mathrm{o}$).}
	\label{delta}
\end{figure}

\begin{figure}[t]
	\vspace*{-0.5cm}
	\hspace*{-0.3cm}
	\centering
	\subfloat[\hspace*{-0.7cm}]{
		\label{previous_approach2}
		\includegraphics[width=0.23\textwidth]{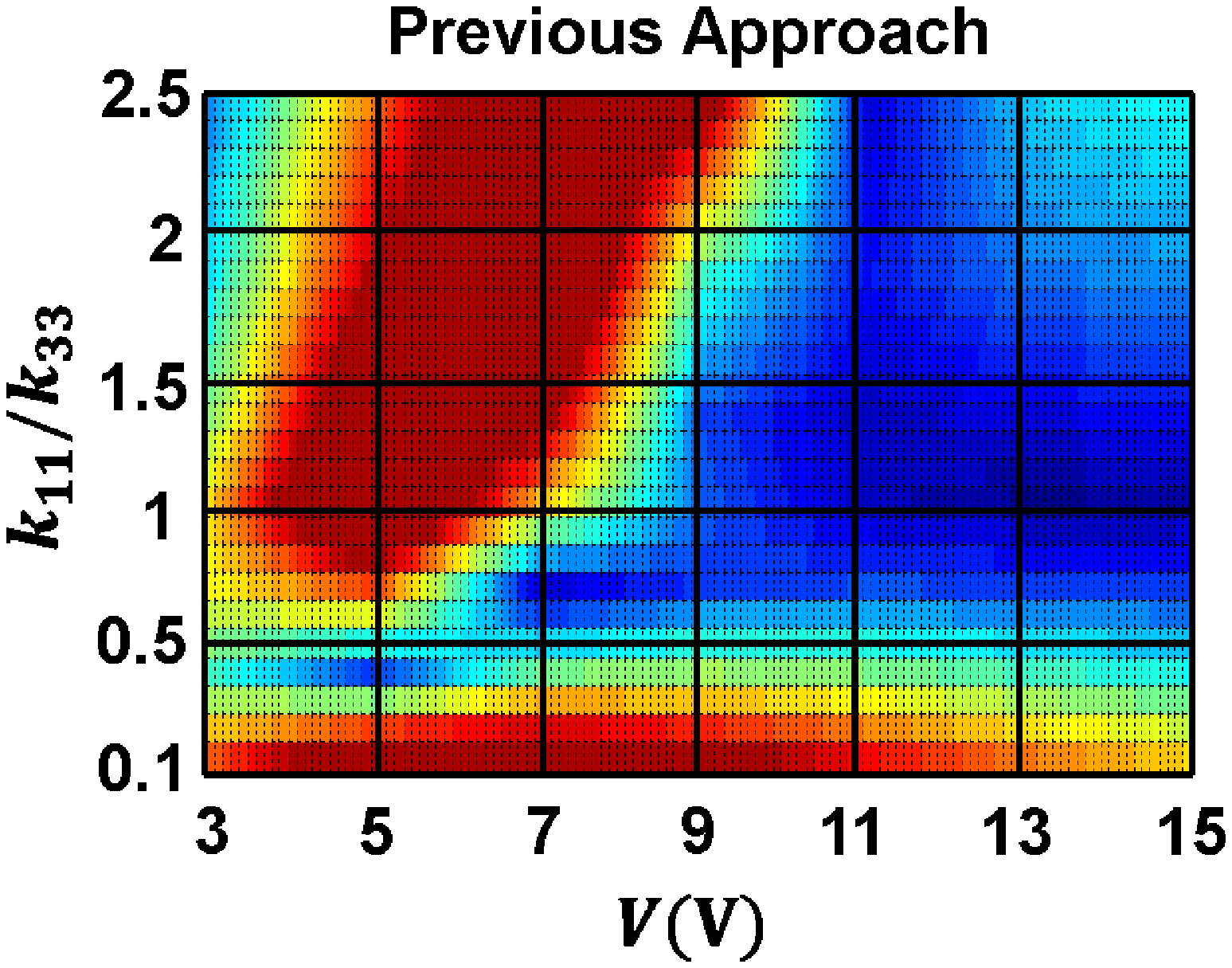}}
	\hspace*{-0.35cm}
	\subfloat[\hspace*{-0.7cm}]{
		\label{integral_approach2}
		\includegraphics[width=0.23\textwidth]{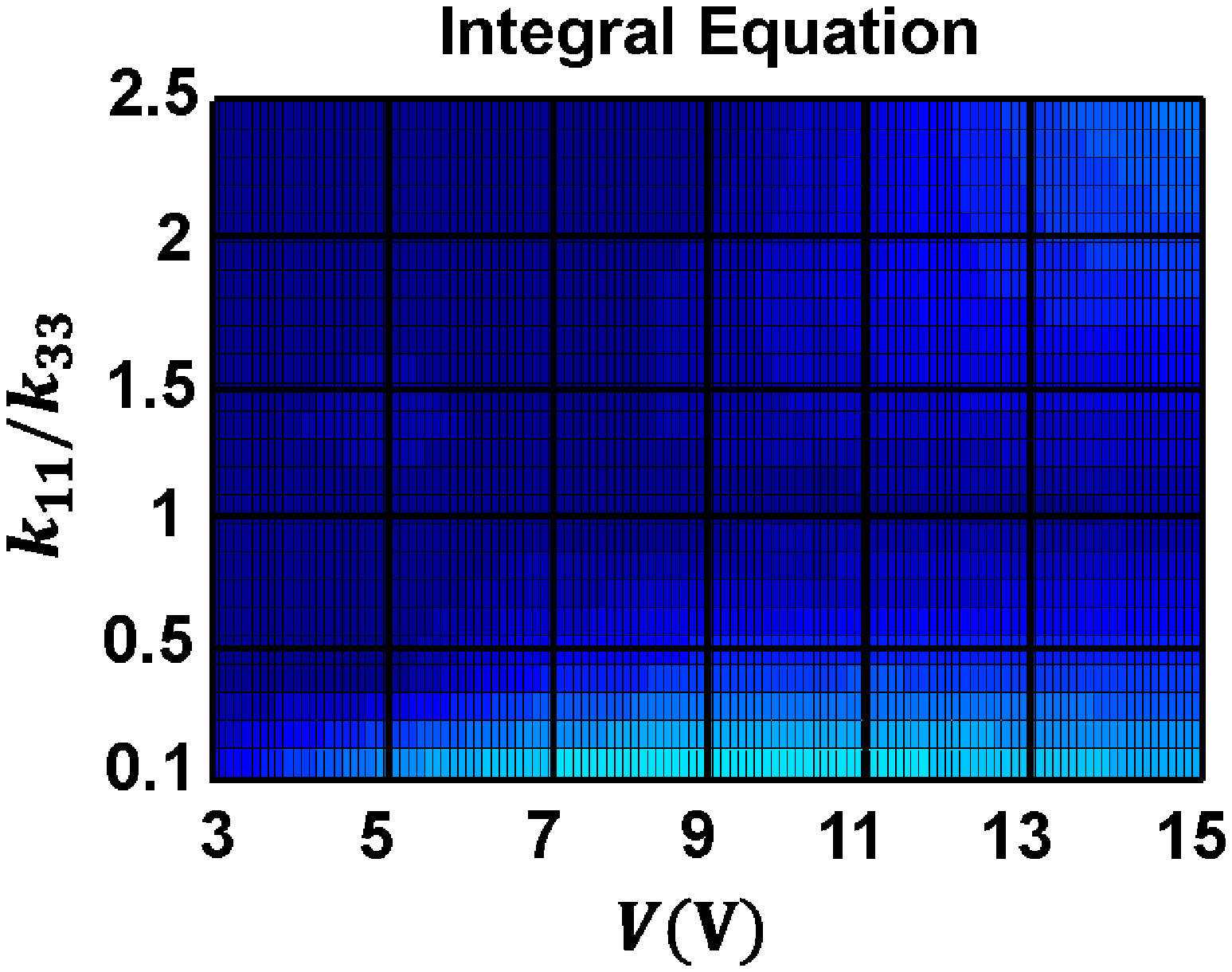}}
	\hspace*{-0.50cm}
	\includegraphics[width=0.059\textwidth]{errorv2}
	\caption{\small Mean absolute error as a function of $V$ and $k_{11}/k_{33}$ obtained with the  approach of \cite{Alex} (a) and the integral-equation formulation (b) ($k_{33}=17.1$ pN, $\Delta \varepsilon=0.47$, $d=200 \, \mu$m, and $\theta_p=4^\mathrm{o}$).}
	\label{k11_k33}
\end{figure}

Figs. \ref{delta} and \ref{k11_k33} present contour plots comparing the mean absolute error obtained with the proposed integral-equation solution and the analytical approach depicted in \cite{Alex}. Numerical estimations are considered as the reference in the absolute error calculation. There are some similarities between Figs. \ref{delta} and \ref{k11_k33}. First of all, the error is considerably reduced in the integral-equation formulation compared to \cite{Alex} for all the cases under study (polarization voltage $V$, dielectric anisotropy $\Delta\varepsilon$, and ratio splay to bend $k_{11}/k_{33}$). Secondly, the error increases drastically in the approach of \cite{Alex} for low polarization voltages, as previously discussed for the 200-$\mu$m Merck E7 LC (Fig. \ref{2D}). In general, this situation is applicable when the parameter $\beta$ is low (either $V$ or $\Delta\varepsilon$ are low).  By contrast, the integral-equation formulation shows a good performance in all the cases. The results presented on E7 are extended to further LC mixtures. In Table 1, the average director is computed with the analytical formula of [12] ($\theta_m^{\textrm{ana}}$), the integral equation ($\theta_m^{\textrm{int}}$) and the numerical solution ($\theta_m^{\textrm{num}}$).


\subsection{Simulation of a Reconfigurable PPW Phase Shifter}
\begin{figure}[!h]
	\hspace*{-0.1cm}
	\includegraphics[width=0.5 \textwidth]{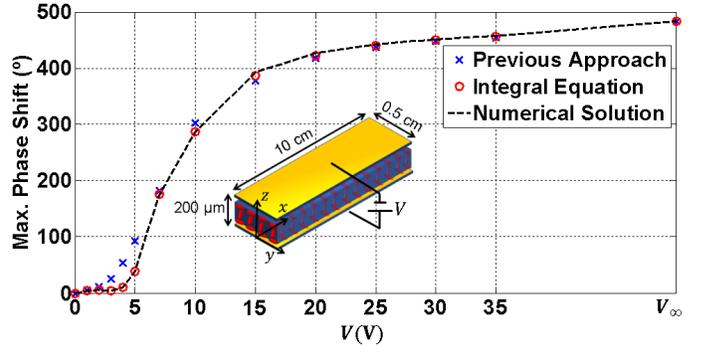}
	\centering
	\caption{\small Maximum phase shift computed in \textit{CST} at 30 GHz in the Merck E7 LC for different polarization voltages.}
	\label{desfase_error}
\end{figure}

 Here, we use the integral equation to estimate the maximum phase shift that can be achieved in a parallel-plate waveguide (PPW) phase shifter filled with a 200-$\mu$m-thick Merck E7 LC \cite{mtt}. Additionally, we compare the accuracy in the estimation with numerical approaches and with the analytical formula of \cite{Alex}. The phase shift is computed using commercial software \textit{CST} through the average director $\theta_m$, and the permittivity tensor that models the liquid crystal \cite{Alex}.  For a 10-cm-long phase shifter at 30 GHz, Fig. \ref{desfase_error} shows the maximum tunable phase shift for every applied voltage $V$. The upper limit is fixed by $V_{\infty}$, which represents a hypothetical infinite voltage. The integral equation fits very well the numerical approach. However, the approach of \cite{Alex} (blue crosses)  presents a  high error  when computing low voltages.

\section{Conclusion}

This paper derives an integral equation that models the director tilting in nematic liquid crystals. The  one-constant approach of \cite{Alex} is generalized for all the values of the elastic constants in RF reconfigurable devices. The agreement between the integral-equation formulation and the numerical solution is remarkable. A deeper study on \cite{Alex} reveals that the analytical approach lacks good results for low values of the polarization voltage and the dielectric anisotropy. By contrast, the accuracy in the integral-equation formulation is preserved. Finally, the integral equation is used to compute via \textit{CST} the phase shift in a PPW phase shifter when varying the polarization voltage. The simulation results based on the integral equation show a high accuracy in the computation.


\begin{table}[!h]
	\centering
	\caption{\small}\vspace{0.1cm}
	\small AVERAGE DIRECTOR ($^\mathrm{o}$) COMPUTED WITH THREE METHODS FOR SOME LC MIXTURES ($\theta_p=4^\mathrm{o}$, $d=200\, \mu$m)
	\label{tab:my-table}\vspace{0.22cm}
	\begin{tabular}{c|c|c|c|}
		\cline{2-4}
		\multicolumn{1}{l|}{}                               & \multicolumn{3}{c|}{\textbf{APPLIED VOLTAGE}}                                                                                                                                                                                                                                                                                                                                                                                                  \\ \hline
		\multicolumn{1}{|c|}{\textbf{LC MIXTURES}}          & \textbf{V=5 V}                                                                                                                                & \textbf{V=10 V}                                                                                                                               & \textbf{V=15 V}                                                                                                                                \\ \hline
		\multicolumn{1}{|c|}{\textbf{E7 {[}15{]}}}          & \begin{tabular}[c]{@{}c@{}}$\theta_m^{\textrm{ana}}=26.16$ \\ $\theta_m^{\textrm{int}}=16.41$ \\ $\theta_m^{\textrm{num}}=16.46$\end{tabular} & \begin{tabular}[c]{@{}c@{}}$\theta_m^{\textrm{ana}}=50.78$ \\ $\theta_m^{\textrm{int}}=52.74$ \\ $\theta_m^{\textrm{num}}=53.37$\end{tabular} & \begin{tabular}[c]{@{}c@{}}$\theta_m^{\textrm{ana}}=63.07$ \\  $\theta_m^{\textrm{int}}=64.48$\\  $\theta_m^{\textrm{num}}=65.38$\end{tabular} \\ \hline
		\multicolumn{1}{|c|}{\textbf{MCL-6608 {[}16{]}}}    & \begin{tabular}[c]{@{}c@{}}$\theta_m^{\textrm{ana}}=16.63$ \\ $\theta_m^{\textrm{int}}=5.68$ \\ $\theta_m^{\textrm{num}}=5.64$\end{tabular}   & \begin{tabular}[c]{@{}c@{}}$\theta_m^{\textrm{ana}}=38.52$ \\ $\theta_m^{\textrm{int}}=32.61$ \\ $\theta_m^{\textrm{num}}=32.55$\end{tabular} & \begin{tabular}[c]{@{}c@{}}$\theta_m^{\textrm{ana}}=52.96$ \\ $\theta_m^{\textrm{int}}=52.73$ \\ $\theta_m^{\textrm{num}}=52.82$\end{tabular}  \\ \hline
		\multicolumn{1}{|c|}{\textbf{GT3-23001 {[}9{]}}}    & \begin{tabular}[c]{@{}c@{}}$\theta_m^{\textrm{ana}}=54.00$ \\ $\theta_m^{\textrm{int}}=55.76$ \\ $\theta_m^{\textrm{num}}=56.40$\end{tabular} & \begin{tabular}[c]{@{}c@{}}$\theta_m^{\textrm{ana}}=71.47$ \\ $\theta_m^{\textrm{int}}=72.31$ \\ $\theta_m^{\textrm{num}}=73.09$\end{tabular} & \begin{tabular}[c]{@{}c@{}}$\theta_m^{\textrm{ana}}=77.50$ \\ $\theta_m^{\textrm{int}}=78.01$ \\ $\theta_m^{\textrm{num}}=78.55$\end{tabular}  \\ \hline
		\multicolumn{1}{|c|}{\textbf{MDA-00-3506 {[}15{]}}} & \begin{tabular}[c]{@{}c@{}}$\theta_m^{\textrm{ana}}=19.52$ \\ $\theta_m^{\textrm{int}}=21.62$ \\ $\theta_m^{\textrm{num}}=21.73$\end{tabular} & \begin{tabular}[c]{@{}c@{}}$\theta_m^{\textrm{ana}}=42.77$ \\ $\theta_m^{\textrm{int}}=49.48$ \\ $\theta_m^{\textrm{num}}=51.31$\end{tabular} & \begin{tabular}[c]{@{}c@{}}$\theta_m^{\textrm{ana}}=56.67$ \\ $\theta_m^{\textrm{int}}=61.20$ \\ $\theta_m^{\textrm{num}}=63.60$\end{tabular}  \\ \hline
		\multicolumn{1}{|c|}{\textbf{MDA-98-1602 {[}11{]}}} & \begin{tabular}[c]{@{}c@{}}$\theta_m^{\textrm{ana}}=64.36$ \\ $\theta_m^{\textrm{int}}=65.59$ \\ $\theta_m^{\textrm{num}}=65.18$\end{tabular} & \begin{tabular}[c]{@{}c@{}}$\theta_m^{\textrm{ana}}=76.94$ \\ $\theta_m^{\textrm{int}}=77.58$ \\ $\theta_m^{\textrm{num}}=77.37$\end{tabular} & \begin{tabular}[c]{@{}c@{}}$\theta_m^{\textrm{ana}}=81.15$ \\ $\theta_m^{\textrm{int}}=81.58$ \\ $\theta_m^{\textrm{num}}=81.41$\end{tabular}  \\ \hline
	\end{tabular}
\end{table}


\end{document}